\documentclass
[showpacs,prl,amsfonts,amssymb,oneside,10pt,preprint,balancelastpage,notitlepage,twocolumn,a4paper]{revtex4}%
\usepackage{amsfonts}
\usepackage{amsmath}
\usepackage{amssymb}
\usepackage{graphicx}%
\providecommand{\U}[1]{\protect\rule{.1in}{.1in}}

\begin{document}
\title{Experimental characterization of domain walls dynamics in a photorefractive oscillator}
\author{Adolfo Esteban--Mart\'{\i}n$^{1}$, Victor B. Taranenko$^{1,2}$, Javier
Garc\'{\i}a$^{1}$, Eugenio Rold\'{a}n$^{1}$, and Germ\'{a}n J. de
Valc\'{a}rcel$^{1}$}
\affiliation{$^{1}$Departament d'\`{O}ptica, Universitat de Val\`{e}ncia, Dr. Moliner 50,
46100--Burjassot, Spain}
\affiliation{$^{2}$Institute of Physics, National Academy of Sciences of the Ukraine, Kiev, Ukraine}

\begin{abstract}
We report the experimental characterization of domain walls dynamics in a
photorefractive resonator in a degenerate four wave mixing configuration. We
show how the non flat profile of the emitted field affects the velocity of
domain walls as well as the variations of intensity and phase gradient during
their motion. We find a clear correlation between these two last quantities
that allows the experimental determination of the chirality that governs the
domain walls dynamics.

\end{abstract}

\pacs{42.65.Sf, 47.54.+r, 42.65.Hw}
\maketitle

\section{Introduction}

Extended nonlinear systems with broken phase invariance (e.g., systems with
only two possible phase values for a given state), are common in nature. These
systems may exhibit different types of patterns but, importantly, the broken
phase invariance lies at the origin of the appearance, in particular, of
domain walls (DWs) which are the interfaces that appear at the boundaries
between two spatial regions occupied by two different phase states
\cite{Manneville,Cross,Walgraef}.

In nonlinear optics there are several examples of spatially extended bistable
systems that can show solutions for the emitted field with a given amplitude
but opposite phases (that is, phases differing by $\pi$), such as degenerate
optical parametric oscillators (DOPOs) or intracavity degenerate four--wave
mixing \cite{Staliunas02}. The interface which connects both solutions, the
DW, can be of either one of two different types: On the one hand, there are
Ising walls in which the light intensity of the field at the core of the
interface is zero and the phase changes abruptly from $\phi$ to $\phi+\pi$; on
the other hand, there are Bloch walls in which the light intensity never
reaches zero and the change of phase is smooth across the DW \cite{Coullet}.
In addition to this, Ising walls are always static whereas Bloch walls are
usually moving fronts (they are static only when the system is variational
what is an uncommon situation for dissipative systems). It is important to
remark that Bloch walls are chiral (they are not identical to their mirror
images) as in the Bloch wall the phase angle rotates continuously through
$\pi$ and two directions of rotation are possible. This fact has important
dynamical consequences as Bloch walls with opposite chirality move in opposite
directions \cite{Coullet}. Both Ising and Bloch walls have been found in
nonlinear optical cavity experiments \cite{Larionova,Esteban05}.

When a control parameter is varied a bifurcation that changes the nature of
the DW may happen. This is the nonequilibrium Ising--Bloch transition (NIBT)
that has been investigated theoretically in \cite{Coullet,deValcarcel02} and
has been repeatidly observed in liquid crystals (see, e.g.,
\cite{Frisch,Kawagishi}). In the context of nonlinear optical systems, the
NIBT has been predicted to occur in type I \cite{Perez04} and type II
\cite{MaxiOL} DOPOs, in intracavity type II second--harmonic generation
\cite{Michaelis}, and in vectorial Kerr cavities \cite{Sanchez05}. Recently,
we have reported the first observation of this phenomenon, the NIBT, in an
optical system, namely a photorefractive oscillator \cite{Esteban05}.
Moreover, our observation is rare in the sense that we observed a hysteretic
NIBT \cite{Esteban05,Taranenko05}.

The aim of the present work is to study in detail the dynamics of the DWs we
reported in \cite{Esteban05} by means of the measurement of the different DWs
characteristics, namely intensity, phase gradient and velocity, establishing
relations among them. In particular, we consider whether the chirality
parameter, which will be described later on, is appropriate for characterizing
the DW.

\section{Experimental setup}

Our experimental setup, Fig.1, is a single-longitudinal mode photorefractive
oscillator (PRO) formed by a Fabry Perot resonator in a near self-imaging
arrangement \cite{Arnaud} chosen in order to achieve a high Fresnel number
\cite{Taranenko98,Esteban04}. The nonlinear material, a \textrm{BaTiO}$_{3}$
crystal, is pumped by two counterpropagating laser beams of the same
frequency. In this way a degenerate four wave mixing process occurs within the
cavity. The degeneracy implies that the field exiting the nonlinear cavity is
phase-locked and only two values of the phase (differing by $\pi$) are allowed
\cite{Yeh}. Hence DW formation is allowed.

The system performance is ruled by different parameters such as detuning
(which is the difference between the frequency of the pump and the frequency
of the cavity longitudinal mode in which emission occurs), gain, losses and
diffraction. All these parameters can be controlled up to some extent. We
choose as in \cite{Esteban05,Esteban04} cavity detuning as the control
parameter as it can be finely tuned in an actively stabilized system
\cite{Vaupel}.

Regarding diffraction, the system is intentionally made quasi--one dimensional
in the transverse dimension (1D system) in order to avoid the influence of DW
curvature in the observed dynamics: Curvature induces a movement in the DW
\cite{Staliunas02,MaxiOL} that contaminates that due to the nature of the DW
(i.e., its Ising or Bloch character). This is achieved by properly placing
slits inside the nonlinear cavity (D in Fig. 1), in particular, at the Fourier
planes (FP in Fig. 1). The width of the slits is adjusted to the size of the
diffraction spot in these planes. In this way beams with too large inclination
(such that their transverse wavevector falls outside the plane defined by the
center line of the slit) are not compatible with the diffraction constraints
of the cavity. This Fourier filtering allows the use of finite width slits and
still gets rid of most 2D effects. It is also by using a diafragm that spatial
frequencies belonging to other longitudinal modes than the one of interest are
removed \cite{Esteban04}.

Detuning, our control parameter, can be changed by means of a piezo-mirror.
Depending on detuning, different types of structures can be found
\cite{Taranenko98,Esteban04,Staliunas97,Mamaev} but for our present purposes
it suffices to say that DWs exist in resonance or for positive cavity detuning
(i.e., when the frequency of the pumping field is smaller than the frequency
of the nearest cavity mode): At zero (or small positive) cavity detuning DWs
are static (Ising type), whilst they start moving (Bloch type) if detuning is
increased enough \cite{Esteban05}.

\section{Injection of Domain Walls}

DWs can form spontaneously from noise when the detuning value lies in the
appropriate domain, as it was the case with the DWs reported in
\cite{Taranenko98,Esteban04}. But waiting for the appearance of DWs from noise
is not the most adequate strategy for their study for several reasons. On the
one hand one must wait until the formation of the DW occurs and it is very
likely that not a single DW but some of them appear, which complicates the
analysis because of DW interaction. On the other hand even when a single
structure is formed from noise, there is the problem that its position on the
transverse plane will not always be the same. Owing to these and other reasons
it is most convenient to find a way for injecting, or writing, DWs at will.

An injection technique to remove the initial structure and write a single DW
has been devised \cite{Esteban05,Esteban05b}. It consists in injecting, for a
short time (the shutter in Fig. 1 remains open for a few seconds), a laser
beam into the photorefractive oscillator that is tilted with respect to the
resonator axis. This tilted beam has a phase profile in the transverse
dimension $x$ given by $\phi\left(  x\right)  =\phi_{0}+\left(  2\pi
/\lambda\sin\alpha\right)  x$, being $\phi_{0}$ a constant, $\lambda$ the
light wavelength, and $\alpha$ the tilt angle with respect to the resonator axis.

Now, this injected tilted beam forces a phase varition, along the transverse
dimension, of the intracavity field phase, but given the phase sensitive
nature of the degenerate four-wave mixing process \cite{Yeh}, only two phase
values (say $0$ and $\pi$, modulo $2\pi$) can be amplified. Thus, the regions
of the field whose phase lies in the domain $\left]  -\pi/2,+\pi/2\right[  $
will be attracted towards the $0$ phase value, whilst those with a phase lying
in the domain $\left]  +\pi/2,+3\pi/2\right[  $ will be attracted towards the
phase value $\pi$.

As an inmediate consequence, the appearance of a DW\ is forced as the points
at which $\phi=\pm\pi/2$ (modulo $2\pi$). These DWs are dark because those
phase values cannnot be amplified but also, and this is a more fundamental
reason, because these points separate adjacent domains with opposite phases.
These points, the DWs, are thus topological defects \cite{Trillo97} what makes
them very robust as they cannot be removed via continous changes.

In Fig. 2 we show several snapshots of the interferogram of the output field
during the writing process. In Fig. 2a the initial homogeneous field is shown,
and the interferogram shows the homogeneity of the field phase. Then, in Fig.
2b the tilted beam is injected. While the injection is being applied, a
blurred DW appears on the left, Fig. 2c, and moves to the right, Figs. 2d-2e.
When the wall is located at the centre of the crystal, Fig. 2f, the injection
is blocked and the wall is generated, Figs. 2g and 2h. In this way a single DW
is written at the desired location along the transverse dimension. Notice that
more than one DW can be written if larger angles in the writing beam are used.

\section{Characterization of the DWs}

In order to fully characterize the features of the different DWs that are
being observed, it is necessary to measure the complex field. This is achieved
by means of the interferometric technique already used in \cite{Larionova}: A
CCD video camera records the interference pattern between the near field
(i.e., the intracavity field at the crystal plane) and a homogeneous reference
beam. From the interferometric pattern, like those of Fig. 2, a Fourier
transform technique allows the reconstruction of both the amplitude and phase
of the optical field, see \cite{Larionova} for full details. Some comments are
in order regarding the way data must be extracted and processed.

An important feature that must be pointed out is that the spatial profile of
the emitted field is not flat, as can be appreciated in Fig. 3. Of course one
cannot expect a perfectly flat profile as the pumping fields are Gaussian, but
the interesting point is that the more positive cavity detuning is the less
flat and less extended the output profile is , i.e., there is a noticeable
self--focusing effect \cite{nota,Vaupel99}. This is particularly relevant in
the "free" transverse dimension ($x$ direction in Fig. 3, which is the
direction that is not limited by the intracavity diafragms).

This fact is at the origin of one effect that has to be taken into account
when characterizing the DW dynamics: The variation of the intensity profile
along the $x$ dimension is quite strong, stronger than for negative detuning,
and no simple experimental arrangement can cope with this. We shall come back
to this point in the following section when studying the dynamics of Bloch
walls as $x$ is the direction along which Bloch walls move.

Apart from these nonuniformities along the $x$ direction, one has to take into
account also the nonuniformities along the direction of the DW, i.e., the $y$
direction, see Fig. 3. The finite extension of the pump as well as the already
commented self--focusing effect make that the output intensity departs
noticeably from a top hat profile. This fact affects the nature of the DW as
it cannot be strictly uniform in the $y$ direction thus making necessary to
discard the borders of the illuminated region in order to characterize DWs properly.

Let us consider, for example, the interferogram corresponding to a static wall
shown in Fig. 4a. For the reconstruction of the field amplitude and phase
first we must choose between taking data for a particular value of $y$ or
making an average for different values of $y$. We choose the second option in
order to smooth the effect of any local imperfection. Then we average the
field obtained from the interferometric technique \cite{Larionova} and obtain
the average field $\left\langle A\left(  x\right)  \right\rangle _{y}$. But
before we must decide if we consider all data or we take into account only the
central part (in the $y$ direction): If only the central part of the
interferogram is considered for reconstructing the field amplitude and phase,
one obtains the profile of an Ising wall, Fig. 4b, with null intensity at the
center and a sharp phase jump. Contrarily, if one does not discard the borders
and takes all data into account, one obtains a profile that does not
correspond so clearly to an Ising wall, as can be seen in Fig. 4c (the
intensity does not reach zero and the phase jump is smoother). Of course this
is nothing but the influence of a border effect that is due to the already
commented nonuniformity of the fields along the $y$ direction, and indicates
the way in which data must be treated because the Ising character of the DW is
clear from its dynamics: They are static.

In the case of moving walls the border effect is not so noticeable: Fig. 5a
shows a typical interferogram of a moving wall and the corresponding averages
of the reconstructed field (without borders in Fig. 5b and with them in Fig.
5c). In this case the wall moves from right to left and it is interesting to
notice that the field amplitude on the left side of the Bloch wall is smaller
than that of the right side. This occurs always: the intensity is larger on
the back side of the wall with respect the direction of movement, a feature
that is reproduced with the simple model used in \cite{Taranenko05} for
reproducing a hysteretic NIBT.

Then the experimental procedure consists in injecting a single DW, recording
the interferogram, extracting the information about intensity, phase gradient
and spatial position of a single DW for different instants of time, and
repeating this procedure for different values of the cavity detuning.

\section{Dynamics of DWs}

In this section we deal with the dynamics of DWs in our particular non flat
background profile. Fig. 6 shows the trajectories followed by both a static
wall and a moving one that were created at the same spatial position (the
centre of the nonlinear crystal). As commented below, the static walls shows
some slow dynamics.

The static wall is an Ising wall, as its intensity and phase profiles for
different intants of time, shown in Fig. 7, clearly show. One can observe that
after an initial transient in which the intensity does not reach zero and the
phase jumps are relatively smooth, the intensity and phase profiles
progresively approach those of an Ising-type: After the transient, the change
of phase through the interface is quite abrupt and the intensity at the center
of the wall is very close to zero. It must also be noticed that the amplitude
of the field on the right and on the left of the DW are similar. The initial
movement can be easily attributed to the transiet behaviour that occurs until
the DW reaches its final profile. The remaining movement must be attributed to
noise as the trajectory shows more a wandering behaviour than a systematic
one. The above completely characterizes static DWs that are clearly identified
as Ising DWs.

Let us now consider moving walls. In Fig. 8 we display their intensity and
phase profiles, again for different time values. The behaviour is very
different from that of the static wall of Fig. 7:\ The moving DW moves
continuously to the left, its velocity changes with time, and the intensity
and phase also change during the motion. We can conclude that it is a Bloch
wall (smooth change of phase and nonzero intensity at the core of the wall).
The fact that the velocity is not constant can be attributed to the fact that
for the different positions the DW is placed on a different background as the
intensity decreases towards the borders of the illuminated region, Fig. 3. In
other words, the acceleration shown by the DW can be attributed to the
existence of an intensity gradient in the field. It is also to be noted that
the intensity in the left side of the DW is smaller than that in the right
side, and also that this effect increases as the wall approaches the border of
the illuminated region. We have checked that all these features are captured
by a simple model for DW dynamics (a Ginzburg--Landau equation with broken
phase symmetry, see \cite{Coullet}) when pump is spatially inhomogeneous.

Looking at how intensity and phase change during the accelerated motion, one
can qualitatively conclude that the larger the wall velocity is, the larger
(smaller) the intensity (the phase gradient) at the centre of the wall is.
These facts make it difficult the determination of the DW velocity for a given
set of parameters: Our procedure has consisted in taking an average velocity
in those cases in which it is more or less constant, discarding always the
velocities corresponding to the regions located close to the border of the
illuminated region (where the acceleration is more obvious).

The measurements we have just commented correspond to a particular value of
the cavity detuning. In order to see how a change in this parameter affects
the moving DW, we show in Fig. 9 the intensity and phase profiles for three
different values of the cavity detuning but corresponding to approximately the
same position on the transverse direction, i.e., to the same intensity
background. We observe again that for increasing detuning the velocity
increases and, again, the more velocity the wall gets, the more intensity and
less phase gradient it has.

To sum up, we can conclude that there is experimental evidence of the
connection between the DW velocity and the characteristics of the optical
field at the centre of the interface: The smoother the phase jump is and the
larger the intensity at the wall core is, the faster the DW is.

\section{Chirality}

In the previous section we have shown how the intensity and phase gradient of
the DW\ change with detuning and background conditions. We pass now to a more
quantitative study of the relation existing between the intensity at the core
of the DW, $I_{0}$, and the phase gradient, $\nabla\phi_{0}$.

In Fig. 10 we represent, in a log--log plot, $\nabla\phi_{0}$ versus $I_{0}$
for fixed parameters, the different points correponding to different time
instants (it is a parametric plot). It is clear that a linear fit is well
suited, which leads to the relation
\begin{equation}
I_{0}^{m}\nabla\phi_{0}=10^{-n},
\end{equation}
where the typical values for the slope and crossing point are $m=-0.56$ and
$n=2.12$, respectively, and the coefficient of determination R--squared has a
value of $0.9989$. A similarly good correlation is also found for other
detuning values, although $m$ and $n$ change slightly with detuning\emph{
}($m$ varies between $-0.67$ and $-0.56$, and $n$ varies between $2.32$ and
$2.12$). It is obvious that these fits do not allow saying that $m$ and $n$
are constant, but we think that their change with detuning could be attributed
to a correponding change with detuning in the shape and intensity of the
background (due to the already commented self--focusing effect). In any case,
this result invites to try to describe the dynamics in terms of only one
magnitude of the field at the interface, either the phase gradient or the
intensity or a combination of both.

As already commented, Bloch walls are chiral structures as the complex field
can rotate in the complex plane when passing from one side to the other of the
DW with two different directions \cite{Coullet}. This has an important
physical consequence, as Bloch walls with different chiralities move in
opposite directions \cite{Coullet}. In \cite{deValcarcel02} a chirality
parameter defined as%
\begin{equation}
\chi=I_{0}\nabla\phi_{0},\label{chi}%
\end{equation}
was used for characterizing the DW dynamics (other definitions of the
chirality can be proposed). For Ising walls the chirality parameter is zero
and the wall remains static, whilst for Bloch walls the chirality is non zero
and the wall moves with a velocity proportional to the value of $\chi$.

From our experimental results we can now justify the use of (\ref{chi}) as a
quantity that characterizes the DW dynamics. In Fig. 11 a linear fit between
the velocity and the chirality parameter of a DW, as it evolves with time, is
shown for a particular value of the detuning. It is worth stressing that the
measurement of the velocities is not so reliable as the measurement of
intensity and phase, for the reasons commented above, and for this reason the
linear fit is not so good as that of Fig. 10. We have also tried using other
quantities as the chirality, the results in Fig. 11 being the best (e.g., if
one uses the intensity or the inverse of the phase gradient as chirality
parameters, the correlations one obtains are $0.9540$ and $0.9813$,
respectively, which are considerably smaller than the $0.9967$ obtained in
Fig. 11).

\section{The Ising--Bloch transition}

There remains to explain how the transition from Ising walls to Bloch walls
(the nonequilibriium Ising--Bloch transition, NIBT) occurs as cavity detuning
is increased. In fact this was the subject of Ref. \cite{Esteban05}. We
reproduce here a brief explanation of this for the sake of completeness.

In Fig. 12 a typical plot of the DW velocity as a function of cavity detuning
is shown. The rare aspect of this NIBT is that it exhibits hysteresis: We
start with a static wall at resonance and increase detuning until the DW
starts moving, signaling an Ising--Bloch transition (this is marked as IBT in
Fig. 12). For larger detuning values, DWs always move. Then, from a large
positive detuning, we inject a DW and check whether it moves or not, and
repite this operation for decreasing detuning values until we obtain an static
(Ising) wall (this is marked as BIT in Fig. 12). The interesting thing is that
the two NIBTs occur at different detuning values. In other words, there is a
finite detuning range (between BIT and IBT) where, for fixed detuning, we
obtain either Ising or Bloch walls in subsequent injections.

Notice that the bistability range an the  velocity values in Fig. 12 are
smaller, roughly by a factor of 50\%, than those reported in \cite{Esteban05}.
The reason is that the two cases correspond to two different experiments with
different gains what obviously affects the DW dynamics quantitatively.

Let us finally comment that a simple model that predicts a hysteretic NIBT
like the one we have just commented can be found in \cite{Taranenko05}. In
this model one finds that the origin of the hysteresys lies on the existence
of bistability in the homogeneous solutions of the system, a bistabillity that
we have also found in the experiment. We refer the interested reader to
\cite{Esteban05,Taranenko05} for further details.

\section{Conclusions}

In conclusion, we have reported a detailed experimental characterization of
domain walls dynamics in a nonlinear optical cavity, namely a photorefractive
oscillator working in a degenerate four--wave mixing configuration. An
injection technique to remove any spontaneous initial structure and to write a
single DW at any desired location has been shown. The dynamics of the DWs has
been characterized in terms of the intensity and phase gradient of the optical
field. We have also shown the relation existing between these two quantities
that leads naturally to the identification of the chirality parameter.
Finally, the hysteretic nonequilibrium Ising--Bloch transition, which turns
out to be hysteretic in our case, has been shown.

\textbf{Acknowledgements}

This work has been financially supported by the Spanish Ministerio de Ciencia
y Tecnolog\'{\i}a and European Union FEDER (Projects BFM2002-04369-C04-01,
FIS2004-06947-C02-01 and FIS2005-07931-C03-01), and by the Ag\`{e}ncia
Valenciana de Ci\`{e}ncia i Tecnologia of the Valencian Government (Project
GRUPOS03/117). V.B.T. was financially supported by the Spanish Ministerio de
Educaci\'{o}n, Cultura y Deporte (grant SAB2002-0240).

{\LARGE FIGURE CAPTIONS}

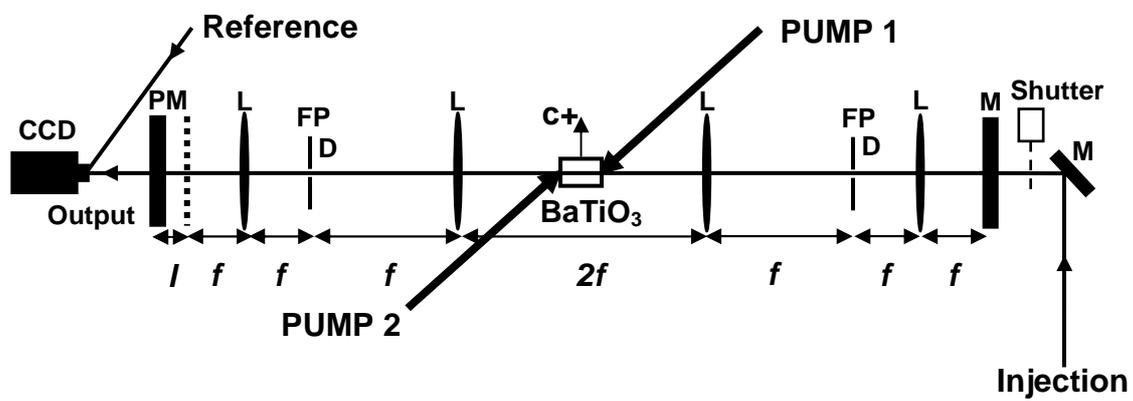
\textbf{Fig. 1.-} Scheme of the experimental setup. M: cavity mirrors; l:
effective cavity length; D: diaphragm located in the Fourier plane (FP) that
makes the system quasi-1D; PM: piezo--mirror for control of the cavity
detuning; CCD: cameras to take pictures from near field; and L: lenses.

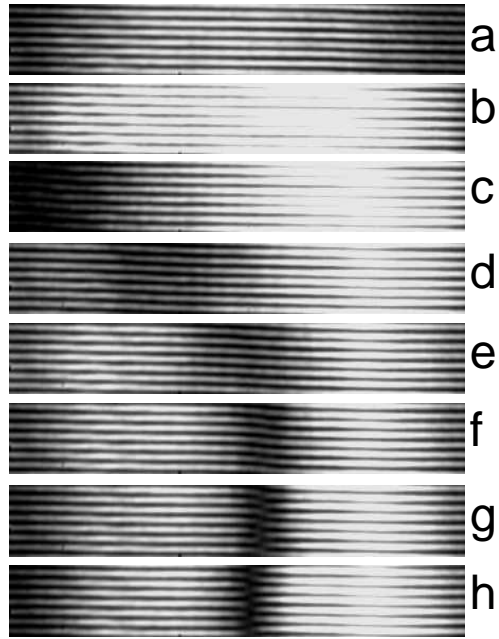
\textbf{Fig. 2.-} Experimental interferometric snapshots of the domain wall
injection process. a) Initial homogeneous field, b) injection is applied,
c)-e) a domain wall appears and moves, f) the injected wall is located at the
centre of the crystal, g) the injection beam is blocked, and h) the domain
wall is finally written. Time runs from top to bottom in steps of 5
\textrm{s}. The transverse dimension is 1.6 mm.

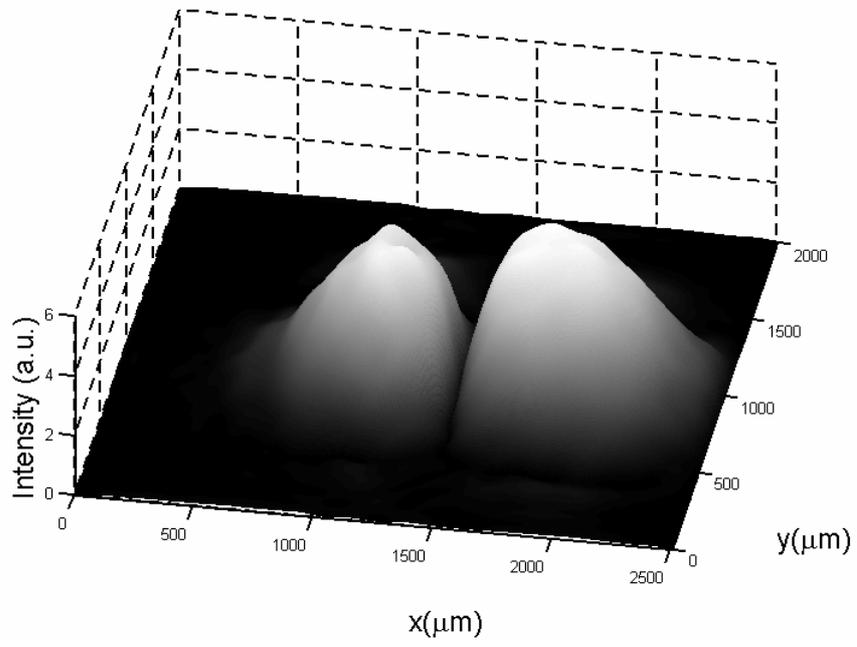
\textbf{Fig. 3.-} Experimental snapshot of a 3D view of the emitted field
containing a domain wall.

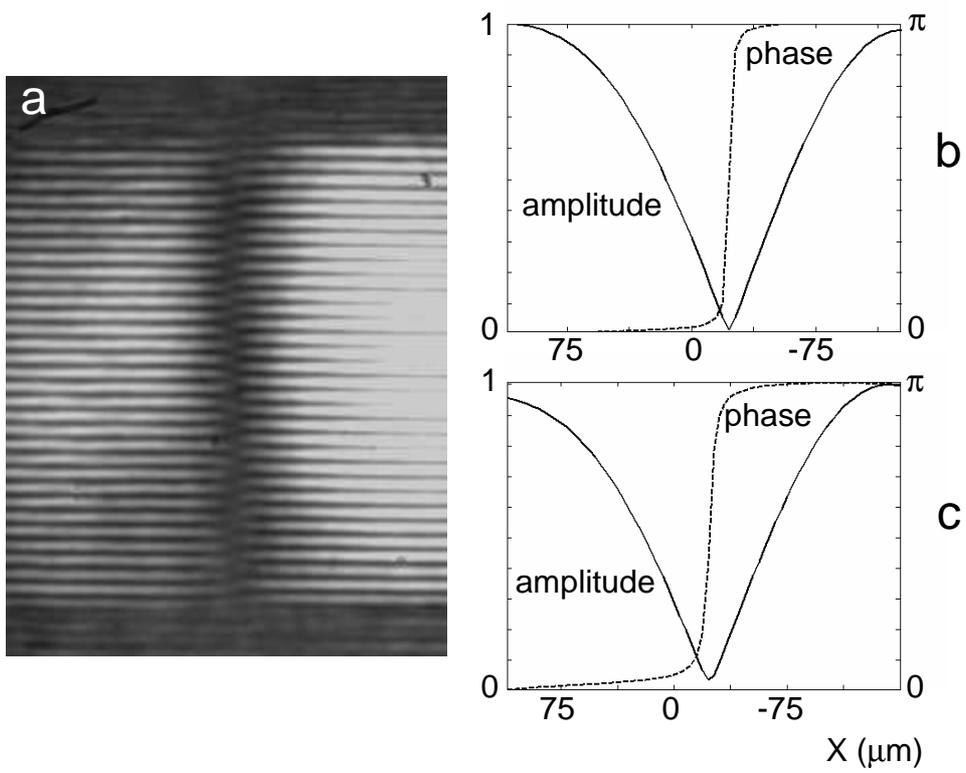
\textbf{Fig. 4.-} Experimental interferometric snapshot of a static domain
wall (a), reconstructed amplitude and phase neglecting the boundaries (b), and
reconstructed amplitude and phase taking into account the boundaries (c). The
cavity detuning is zero.

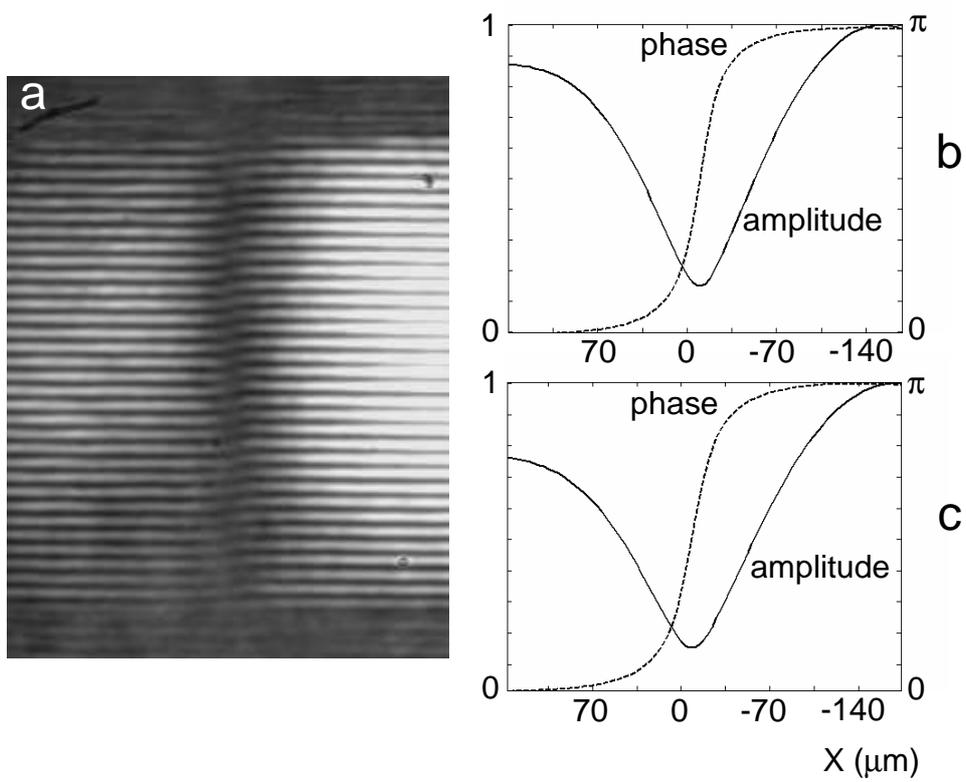
\textbf{Fig. 5.-} Experimental interferometric snapshot of a static domain
wall (a), reconstructed amplitude and phase neglecting the boundaries (b), and
reconstructed amplitude and phase taking into account the boundaries (c). The
cavity detuning is 12\% of the free spectral range (FSR). The FSR of the
cavity is 120Mhz.

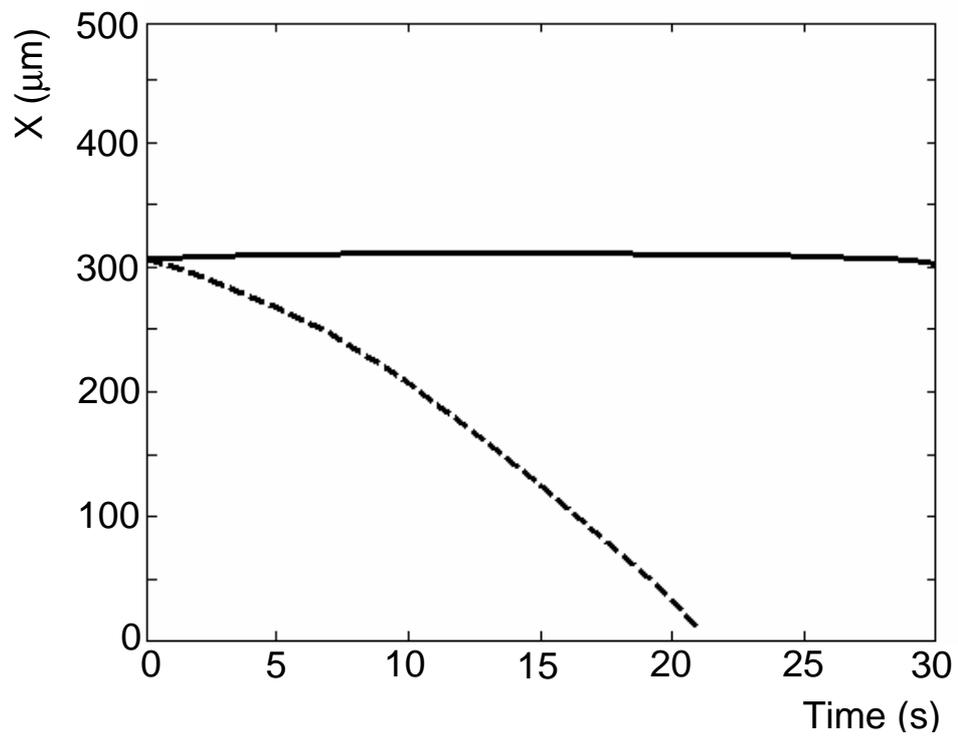
\textbf{Fig. 6.-} Plot of the trajectories of a static (full line) and a
moving (dashed line) domain wall that were generated at the same spatial
position. The detuning for the static and moving wall are zero and 20\% of the
free spectral range, respectively.

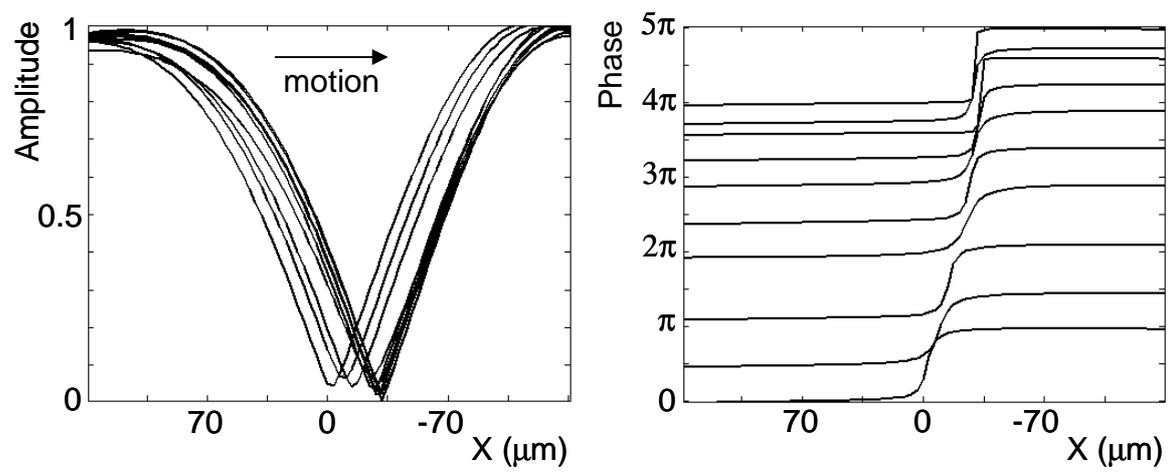
\textbf{Fig. 7.-} Reconstructed amplitudes (top) and phases (bottom)
corresponding to a static domain wall at different instants (the time interval
between different profiles is 5s). The different location of the phase
gradients (bottom) is artificial and has been introduced for the sake of
clarity (this applies also to Fig. 8).

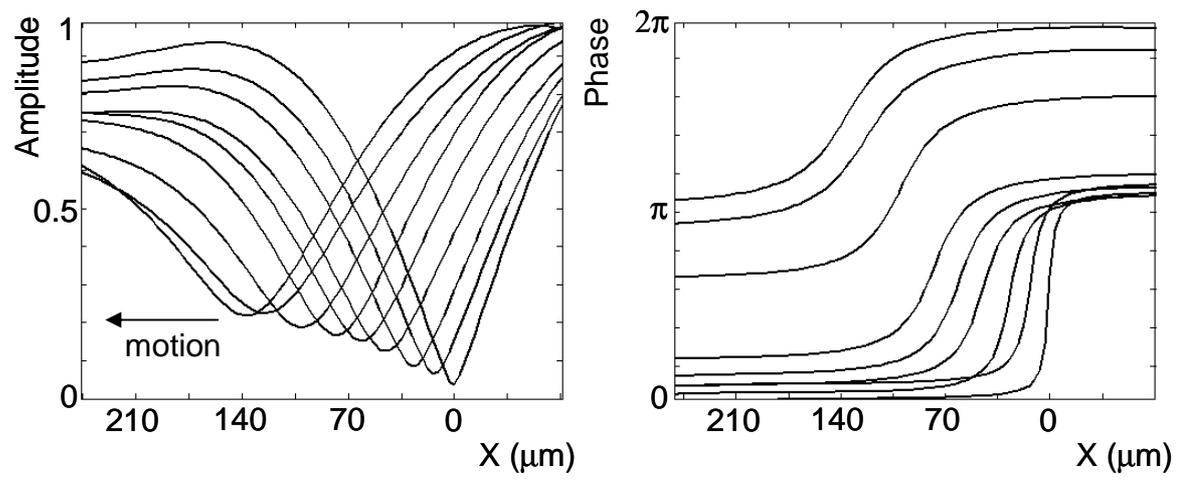
\textbf{Fig. 8.-} As Fig. 7 but for a moving domain wall. The time interval
between different profiles is 2s.

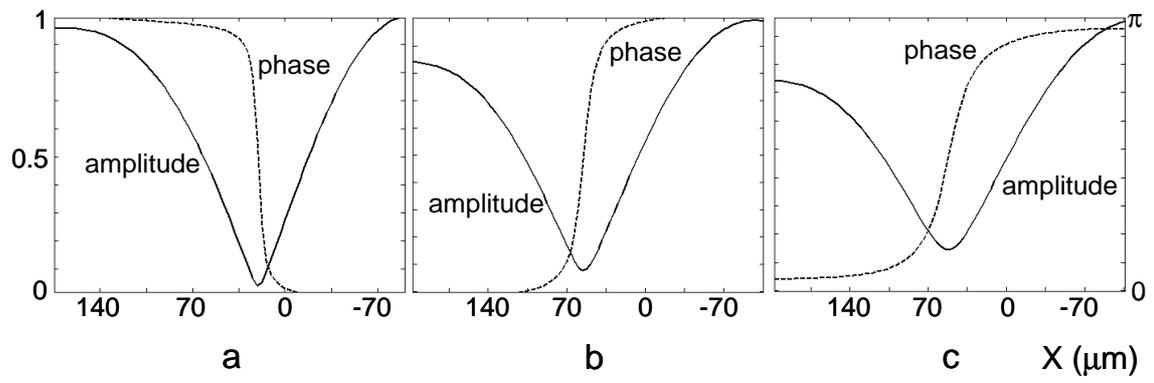
\textbf{Fig. 9.-} Reconstructed amplitudes and phases of domain walls located
at the same spatial position but obtained for different detunings. Detuning
increases from left to rigth (in percentage of the free--spectral range): a) 0
\% , b) 10 \% and c) 20 \%. Velocity increases as detuning does: a) $0$
$\mathrm{\mu m}.\mathrm{s}^{-1}$, b) $7$ $\mathrm{\mu m}.\mathrm{s}^{-1}$, and
c) $12$ $\mathrm{\mu m}.\mathrm{s}^{-1}$.

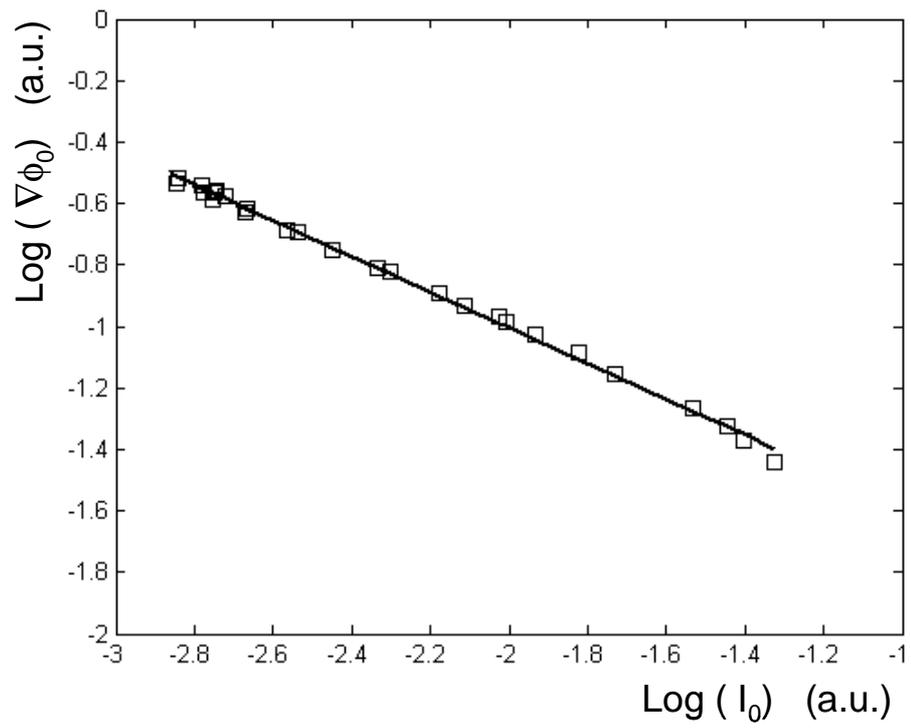
\textbf{Fig. 10.-} Logarithm of the intensity at the core of the DW, $I_{0}$,
versus the logarithm of the phase gradient, $\nabla\phi_{0}$, for different
instants and fixed detuning (20\% of the free--spectral range). Squares
correspond to experimental data and the straight line correspond to the fit.

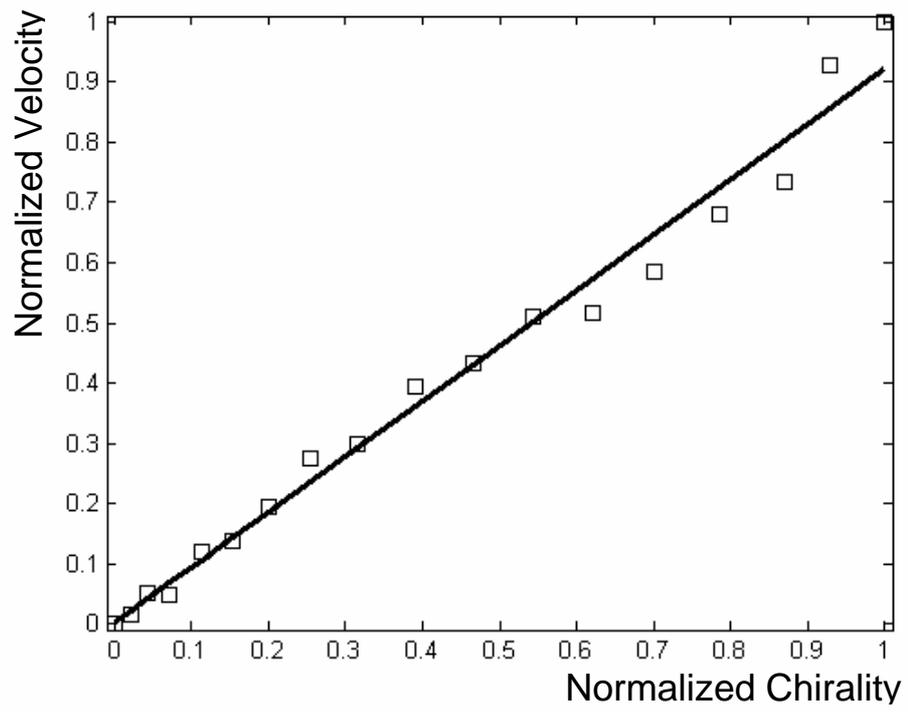
\textbf{Fig. 11.-} Relation between the velocity of a moving domain wall and
the chirality parameter measured during its motion in a non flat background
field. Squares correspond to the experimental data and straight lines
correspond to the linear fit.

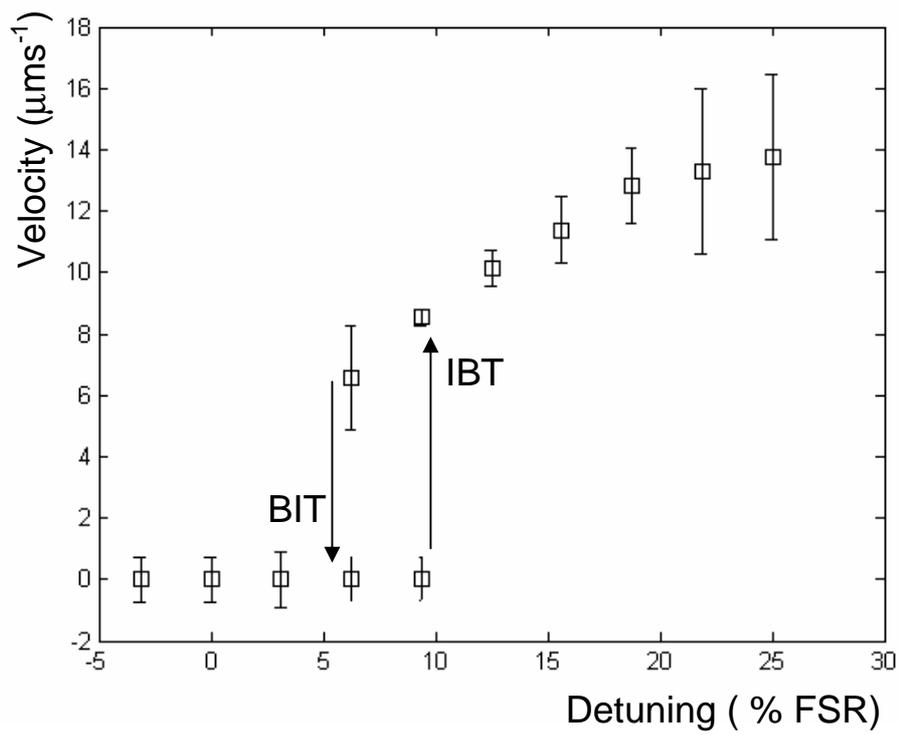
\textbf{Fig. 12.-} Velocity of the DWs as a function of cavity detuning. BIT
and IBT mark the two nonequilibrium Ising--Bloch transitions. See text for
more details.


\begin{thebibliography}{99}
\bibitem {nota}We note that for negative cavity detuning (where a stripe
pattern is observed, see \cite{Esteban04}) a wide output beam is obtained in
the $x$ direction. As detuning is made positive and increased the output beam
becomes increasingly narrower in the "free" transverse dimension. This
observation is compatible with the observations in \cite{Vaupel99}.
\end{thebibliography}
\end{document}